\documentclass[aps,showpacs]{revtex4}

\RequirePackage{amsfonts,amssymb,amsmath}
\usepackage{epsfig}

\begin{document}
\title{ Quantum Singularities in Spacetimes with Spherical and Cylindrical Topological Defects.}
\author{Paulo M. Pitelli} 
\email{e-mail:pitelli@ime.unicamp.br}
\author{Patricio S. Letelier} 
 \email{e-mail: letelier@ime.unicamp.br} 
\affiliation{
Departamento de Matem\'atica Aplicada-IMECC,
Universidade Estadual de Campinas,
13081-970 Campinas,  S.P., Brazil}

\begin{abstract}

Exact solutions of Einstein equations with null Riemman-Christoffel curvature 
tensor everywhere,  except on a hypersurface, are studied using quantum particles
obeying the Klein-Gordon equation. We consider the particular cases when the
 curvature is  represented by a Dirac delta function with support either  on a 
sphere or on  a cylinder (spherical and cylindrical shells). In particular, we 
analyze the necessity of  extra boundary conditions on  the shells.

\end{abstract}

\maketitle

\section{Introduction}

Topological defects appear naturally in theories of the early universe based 
on the spontaneous symmetry breaking of some unifying group. Examples are  the
 cosmic string, which appears in the breaking of a $U(1)$ symmetry group and the 
cosmic wall, which is produced in the breaking of a discrete symmetry. These
 defects are characterized by a null curvature tensor  everywhere, except on a 
submanifold, where it is  proportional to a Dirac delta function.

These spaces are classically singular, Wald \cite{wald}, followed by Horowitz 
and Marolf \cite{horowitz}, establishes an analogy between the classical  and 
quantum singularities. The trajectories of classical particles (geodesics)
used to test classical singularities are replaced by solutions to a
quantum mechanics equation. Spacetimes with a $0$-dimensional singularity,  as 
well as,  spacetimes with  a $1$-dimensional singularity  have been classically
 and quantum mechanically studied \cite{wald}-\cite{konkowski3}.

 In a seminal  paper on this subject, Wald \cite{wald} considered an arbitrary
 self-adjoint extension of the spatial portion of wave operator in an arbitrary 
static spacetime (with singularities consistent with staticity) and showed that
 the resulting solution agreed with the usual Cauchy evolution inside the domain
 of dependence of the initial surface. 

Horowitz and Marolf  \cite{horowitz}  considered a spacetime as quantum 
mechanically nonsingular when the evolution of a general state is uniquely
 defined for all time. They showed that there exist a general class of
 spacetimes with naked singularities  which are nonsigular quantum mechanically. 

Ishibashi and Hosoya \cite{ishibashi} studied several spacetimes with naked 
singularities using Sobolev spaces (instead of $L^2$) as the function space of 
initial data. They showed that some spacetimes, which remain quantum
 mechanically singular with a $L^2$  function space, become regular when probed
 by waves in Sobolev space (for example, the negative mass Schwarzchild 
spacetime). But, the Sobolev space is not the natural space in quantum mechanics. 

Kay and Studer \cite{kay} studied the $2$-dimensional cone (the spacetime 
generated by a point source in a $3$-dimensional Einstein gravity without 
cosmological constant \cite{jackiw}), described by the metric
\begin{equation} 
ds^2=-dt^2+dr^2+\beta^2r^2d\theta^2,
\end{equation}
where the constant $ \beta$ is related to the mass of the  source. They showed 
that this spacetime remains singular when tested by quantum particles and found 
all the possible self-adjoint extensions necessary to turn self-adjoint the
 spatial portion of the wave operator. They are represented by a non trivial 
$1$-parameter family of boundary conditions.

Helliwell, Konkowski and Arndt \cite{konkowski2} studied a $1+3$ dimensional 
spacetime with a cosmic dislocation and a disclination  at $z=0$  given by 
the metric \cite{galet},
\begin{equation}
ds^2=-dt^2+dr^2+\beta^2r^2d\theta^2+(dz+\gamma  d\theta)^2,
\end{equation}
where $\gamma$ is a screw dislocation parameter. They found that this spacetime
 remains singular and generalized the results to Maxwell and Dirac fields, with 
the same results.

To the best of our knowledge  the spacetime singularities for defects of 
dimensions grater than one have not been studied from a quantum mechanical view 
point. The aim of this  work  is to study the evolution of quantum particles 
in spacetimes with defects on hypersurfaces, specifically on spherical (bubbles)
 and cylindrical shells. 

This paper is organized as follows. In section \ref{section 2} we discuss some 
classical aspects of spacetimes with spherical and cylindrical walls. In section
 \ref{section 3} we present a short review of quantum singularities in a general
 static spacetime, based mainly on references \cite{wald} and \cite{horowitz}. 
In section \ref{section 4} we study the quantum behavior of particles in 
spacetimes with spherical and cylindrical defects and discuss about necessity 
of extra boundary conditions on the classical singular wall. In section 
\ref{section 5} we find a $1$-parameter family of boundary conditions for 
spherical, as well as, cylindrical shells. In section \ref{section 6} we give a 
simple example which illustrates the results of the two previous sections. 
Finally, in section \ref{section 7}, we discuss some of our  results  and point 
out  generalizations to  spacetimes with shells with  other symmetries  and to 
quantum particles obeying Maxwell and Dirac equations.


\section{Spherical and Cylindrical Walls}
\label{section 2}

Spacetimes with a topological defect on a general hypersurface $\phi=0$ can be 
constructed splitting the space in two regions, say  $\Omega^{+}$ and
 $\Omega^{-}$, where $\phi>0$ and $\phi<0$,  respectively. By the requirement 
that the Riemann-Christoffel tensor, $R_{\mu\nu\sigma\eta},$  be null on
 $\Omega^{+}$ and $\Omega^{-}$, we obtain a spacetime similar to
 Minkowski spacetime, except for the defect. 

We demand  continuity of  the  metric   and  discontinuity of its first
 derivatives on $\phi=0$. The dependence of $R_{\mu\nu\sigma\eta}$ on the
 second derivative of 
$g_{\mu\nu}$ gives us a delta function  with support on the defect. 

Spherical and cylindrical defects  are obtained  with the function
 \mbox{$\phi=r-r_{0}$}, where $r$ is the radial coordinate on both cases. In 
the spherical case metric is \cite{letelier1}, 
\begin{equation}
\begin{aligned}
&ds^2=-dt^2+dr^2+(r-b)^2d\Omega^2,\; &r>r_{0}\\
&ds^2=-dt^2+dr^2+(r-a)^2d\Omega^2,\; &r<r_{0},
\label{esfera}
\end{aligned}
\end{equation}
where $d\Omega^2$ is the usual measure on $\mathbb{S}^2$, 
\mbox{$b=r_{0}+4/\rho_{0}$} and $a=r_{0}-4/\rho_0$, $\rho_{0}$ is a positive 
constant. This metric represents a static spherical shell of radius
 \mbox{$r_{0}=(a+b)/2$}, centered at the origin, with density 
\mbox{$\rho=\rho_{0}\delta(r-r_{0})$}. Note that on $r=r_{0}$, the metric is
 continuous, while its first derivatives have a discontinuity given by
\begin{equation}
\begin{aligned}
&[g_{\theta \theta,r}]=\lim_{r \to r_{0}^{+}}{g_{\theta \theta,r}}-
\lim_{r \to r_{0}^{-}}{g_{\theta \theta,r}}=2(a-b),\\
&[g_{\phi \phi,r}]=\lim_{r \to r_{0}^{+}}{g_{\phi \phi,r}}-\lim_{r 
\to r_{0}^{-}}{g_{\phi \phi,r}}=2(a-b)\sin^2{\theta}.
\end{aligned}
\end{equation}

The non-null components of curvature tensor $R_{\mu\nu\lambda\sigma}$, viewed as
 a distribution \cite{letelier2}, are
\begin{equation}
\begin{aligned}
&R_{1212}=(b-a)\delta(r-r_0)\\
&R_{1313}=(b-a)\sin^2{\theta}\delta (r-r_0).
\end{aligned}
\end{equation}
The non-null components of the energy-momentum tensor are
\begin{equation}
\begin{aligned}
&T_{00}=(b-a)\delta (r-r_0)\\
&T_{11}=-(b-a)\delta (r-r_0).
\end{aligned}
\end{equation}

Similarly, for the cylindrical case, we have
\begin{equation}
\begin{aligned}
&ds^2=-dt^2+dr^2+(r-b)^2d\varphi^2+dz^2,\; & r>r_{0}\\
&ds^2=-dt^2+dr^2+(r-a)^2d\varphi^2+dz^2,\; & r<r_{0}
\end{aligned}
\label{cilindro}
\end{equation}
but now $b=r_{0}+2/\rho_{0}$ and $a=r_0-2/\rho_{0}$. The metric in this case
 represents a static cylindrical shell centered at the $z$ axis, of radius
 $r_{0}=(a+b)/2$ and density $\rho=\rho_{0}\delta(r-r_{0})$.

Note that on both cases, metrics inside and outside the shells, are  obtained
 from the Minkowski spacetime by an isometric coordinate transformation
 $r\longrightarrow r-\textrm{constant}$. Hence, they are isometric to Minkowski spacetime.

Note that if one allows the radial coordinate take negative values  we have
 two Minkowski spacetimes separated by a defect, i.e., a wormhole, which 
makes these spacetimes particularly interesting.

The points $r=a$ and $r=b$ are metric singularities of the polar type, without
 any deeper geometrical meaning. The real spacetime singularity is on 
 $r=r_{0}$, the middle point between $a$ and $b$.

Figure $1$ shows timelike geodesics on the external side of the bubble
 \cite{letelier1}, in the plane $\theta=\pi/2$. The continuous circle represents
 the physical singularity $r=r_{0}$, whereas the dashed one represents the 
metric singularity $r=b$. The figure shows that the surface $r=b$ is 
completely regular. The geodesics cross this surface with no problem.

\begin{figure}[h]
\centering \epsfig{file=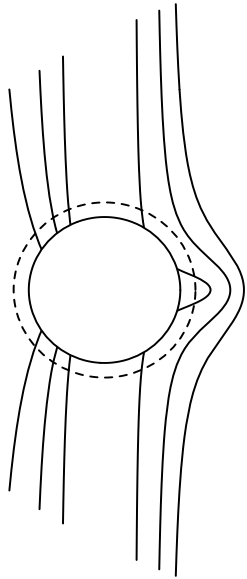, height=8cm,
width=3.3cm} \caption{ {\small Motions of equal energy particles, with distinct
 angular momentum, subjects to the gravitational field generated by a spherical 
defect on the external side of the shell. Some of these geodesics end abruptly 
on the surface $r=r_{0}$, while the surface $r=b$ is perfectly regular.}}
\end{figure}

   
\section{Quantum Singularities}
\label{section 3}

In general relativity, spacetime singularities are indicated by incomplete 
geodesics. At points where a geodesic abruptly ends, extra information is needed,
 since it is not possible to predict the future of a free falling particle in 
an incomplete geodesic spacetime. In quantum mechanics, this extra information 
corresponds to a boundary condition at the classical singular points needed to 
turn self-adjoint the spatial part of the wave operator. When this extra 
information is not necessary the evolution of the wave packet is uniquely
 determined by the wave function at $t=0$. We say that space is not quantum 
mechanically singular. In such cases, the condition that the solutions be 
square-integrable works as a boundary condition and the initial conditions 
alone determine time evolution of the particles.

An example of a classical singular theory, which becomes nonsingular in the view
 of quantum mechanics, is the nonrelativistic hydrogen atom. The imposition of
 quadratic integrability of the solutions of Schr\"odinger equation are 
sufficient to provide a complete set of eingenfunctions. Given an initial wave
 packet, its time evolution is uniquely determined.

Another example, now of a singulity that remains singular when tested by 
quantum mechanics, is the nonrelativistic particle trapped in a $1$-dimensional
 box. A boundary condition is necessary on both edges (it is usually taken
$\psi(0)=\psi(1)=0$) in order to evolve uniquely the wave packet.
Now we shall  present the concepts mentioned  previously in a more precise way.

Let $(M,g_{\mu\nu})$ be a static spacetime with a timelike Killing vector 
field $\xi^{\mu}$. Also,  let $t$ be the Killing parameter and $\Sigma$ a static
 spacelike slice (without any singularity)  orthogonal to the Killing vector
 for all $t$. The Klein-Gordon equation on this space is
\begin{equation}
\square\Psi=M^2\Psi,
\label{Klein-Gordon_0}
\end{equation}
where $\square=\nabla^{\mu}\nabla_{\mu}$ can be split in a temporal and 
a spatial part, like
\begin{eqnarray}
\frac{\partial^2 \Psi}{\partial t^2}&=&-A\Psi \\
&=& VD^{i}(VD_{i}\Psi)-M^2 V^2 \Psi,
\label{separada}
\end{eqnarray}
with $V=-\xi^{\mu}\xi_{\mu}$ and $D_{i}$ the spatial covariant derivative
 on the slice $\Sigma$.

To find the domain of the operator $A$, $D(A)$, is a more difficult task and 
generally no information is provided. So, a minimum domain is 
taken (a core), where the operator can be defined and which does not 
enclose the spacetime 
singularities. An appropriate set is $C_{0}^{\infty}(\Sigma)$, the set of the
 smooth function of compact support on $\Sigma$. On the Hilbert space
 $L^2(\Sigma, V^{-1}d\mu)$, where $d\mu$ is the proper element on $\Sigma$, it 
is not difficult to show (integrating by parts) that the operator 
$(A,C_{0}^{\infty}(\Sigma))$ is a positive symmetric operator. Then, 
self-adjoints extensions always exist \cite{reed-simon1} (at least
 the Friedrichs extension).
Let $A_{E}$ be such extension, obtained by relaxing the boundary condition 
so that the extended domain coincides with the domain of its adjoint operator. 
Because of the self-adjointness of $A_{E}$, the time evolution of the field
 $\Psi$ is uniquely determined by the initial wave packet and is given by
\begin{equation}
\Psi(t)=\cos(A_{E}^{1/2}t) \Psi(0)+A_{E}^{1/2}\sin(A_{E}^{1/2}t)\dot{\Psi}(0),
\end{equation}
where functions of the operator $A_{E}$ exist by the 
spectral theorem \cite{reed-simon1}.

For the nonrelativistic case, the situation is similar. The Schr\"odinger
 equation can be written as, 
\begin{equation}
i\frac{\partial \Psi}{\partial t}=-\nabla^2 \Psi,
\end{equation}
where $\nabla^2$ is the Laplace-Beltrami on $(\Sigma, h_{ij})$ and
 $h_{ij}$ is the induced metric on $\Sigma$. If $-\nabla^2_{E}$ is 
a self-adjoint extension of $-\nabla^2$, the evolution of the particle
 is given by
\begin{equation}
\Psi(t)=e^{-i\nabla^{2}_{E}t}\Psi(0).
\end{equation}

If many self-adjoint extensions exist, the choice of one is necessary in order
 to obtain the time evolution of the particle. Extra information is needed 
(boundary conditions), the spacetime is quantum mechanically singular and we 
can clearly see the resemblance to the classical case. However, if there is 
only one self-adjoint extension, the operator $A$ is said essentially 
self-adjoint and the quantum evolution of the particle is uniquely 
determined by the initial packet. The spacetime is quantum 
mechanically nonsingular. 

To  study  quantum singularities of a static spacetime, i.e., 
the essentially self-adjointness of the spatial operator $A$ defined 
in equation (\ref{separada}), a tool  used  is the  von Neumann theorem  
\cite{reed-simon2}, which says that the self-adjoint extensions of a closed 
Hermitian  operator $T$ is in a one-to-one correspondence with the partial 
isometries of $\textrm{Ker}(T^{\ast}-i)$ into $\textrm{Ker}(T^{\ast}+i)$, where 
$\textrm{Ker}(T^{\ast}\pm i)$ denotes the kernel of $T^{\ast}\pm i$ and
  $T^{\ast}$ denotes the Hilbert adjoint operator of $T$.

In our case, the domain of $A$ is so small, i.e., the restrictions on the 
functions in $D(A)=C_{0}^{\infty}(\Sigma)$ are so strong, that the domain of the
 adjoint operator becomes extremely wide and no restrictions on 
$\psi \in D(A^{\ast})$ are necessary, except that $A^{\ast}\psi \in L^2$.
Hence

\begin{equation}
D(A^{\ast})=\{\psi \in L^2: A^{\ast}\psi \in L^2\}.
\label{critério}
\end{equation}

So, we must solve the following equations
\begin{equation}
A^{\ast}\psi \mp i\psi=0 
\label{teste}
\end{equation}
and count the number of solutions in $L^2$.
We define $n_{\pm}=\textrm{{\it dim}}\;\textrm{Ker}(A^{\ast}\mp i)$
 and name them the deficiency indecis \cite{richtmyer,reed-simon1,reed-simon2}.

If do not exist  solutions of (\ref{teste}) in $L^2$, then $n_{+}=n_{-}=0$ and 
$A$ is essentially self-adjoint. If $n_{+}=n_{-}=1$, there is a $1$-parameter 
family of isometries of $\textrm{Ker}(A^{\ast}-i)$ into
 $\textrm{Ker}(A^{\ast}+i)$, hence there is a $1$-parameter  
family of self-adjoint extensions. The case $n_{+}=n_{-}=2$ is similar.


\section{Quantum Mechanics Around Bubbles and Cylindrical Shells}
\label{section 4}

\subsection{Bubbles}

In the spacetime described by the metric (\ref{esfera}), Klein-Gordon equation 
is similar to the wave equation in flat spacetime, except by the isometric 
transformation $r \to r-\textrm{constant}$,
\begin{equation}
\nabla_{\mu}\nabla^{\mu}\Psi=-\frac{\partial^2 \Psi}{\partial t^2}+
\nabla^2\Psi=M^2\Psi.
\label{Klein-Gordon local coordinates}
\end{equation}

In our case
\begin{equation}
\begin{aligned}
&\frac{\partial^2\Psi}{\partial t^2}=\frac{\partial^2\Psi}{\partial r^2}
+\frac{2}{r-b}\frac{\partial \Psi}{\partial r}-\frac{{\bf L}^2}{(r-b)^2}
-M^2\Psi \\&
\frac{\partial^2\Psi}{\partial t^2}=\frac{\partial^2\Psi}{\partial r^2}
+\frac{2}{r-a}\frac{\partial \Psi}{\partial r}-\frac{{\bf L}^2}{(r-a)^2}
-M^2 \Psi,
\end{aligned}
\label{Klein-Gordon}
\end{equation}
where ${\bf L}^2$ is the square of angular momentum operator.

As we said previously, we must solve the equations
\begin{displaymath}
A^{\ast}\psi \mp i\psi=0,
\end{displaymath}
where $A$ is the spatial portion of the wave equation (right hand side
 of equation (\ref{Klein-Gordon})).

The equation separates using $\psi=R(r)Y_{l}^{m}(\theta, \varphi)$, 
where $Y_{l}^{m}(\theta, \varphi)$ are the usual spherical harmonics. The 
radial part is
\begin{equation}
\begin{aligned}
&R_{l,m}''(r)+\frac{2}{(r-b)}R_{l,m}'(r)+\bigg[(\pm i - M^2)
-\frac{l(l+1)}{(r-b)^2}\bigg]R_{l,m}(r)=0\\&
R_{l,m}''(r)+\frac{2}{(r-a)}R_{l,m}'(r)+\bigg[(\pm i - M^2)
-\frac{l(l+1)}{(r-a)^2}\bigg]R_{l,m}(r)=0.
\label{teste radial}
\end{aligned}
\end{equation}

The spacetimes on both sides of the bubble are completely independent 
(they are separated by a wall), so the appropriate Hilbert space is the 
tensorial product of the $L^2$ spaces on each side of the bubble. If we 
call $\Sigma_{1}$ the outer spatial slice and $\Sigma_{2}$ the 
inner one, we have 
\begin{equation}
H=L^2(\Sigma_{1},(r-b)^2\sin{\theta}drd\theta d\varphi)\otimes
 L^2(\Sigma_{2},(r-a)^2\sin{\theta}drd\theta d\varphi).
\end{equation}

Therefore we shall  analyze the spatial portion of the wave operators on each 
side of the bubble. Let us look first to the external side of the bubble, as 
$r$ goes to infinity, the last term in equation (\ref{teste radial}) can be 
neglected and the asymptotic solution is
\begin{equation}   
R(r)=\frac{1}{(r-b)}[C_1e^{\alpha (r-b)}+C_2e^{-\alpha (r-b)}],
\label{solucao infinito}
\end{equation}
where 
\begin{equation}
\alpha=\frac{1}{\sqrt{2}}\bigg[(\sqrt{1+M^4}+M^2)^{1/2} \mp 
i (\sqrt{1+M^4}-M^2)^{1/2}\bigg].
\end{equation}

Obviously, solution (\ref{solucao infinito}) is square-integrable near infinity 
only if $C_1=0$. The behavior near $r=r_{0}$ does not really matter in this
 case, this point is an ordinary point of equation (\ref{teste radial}). Then, 
both solutions are continuous in $r=r_{0}$, hence, square-integrable near 
this point. So we can adjust the constants of the general solution  in order 
to meet the asymptotic behavior at infinity $R(r)\sim(1/r)e^{-\alpha (r-b)}$. 

There is one solution in $L^2$, to each equation in (\ref{teste}). Hence 
$n_{\pm}=1$ and exists a one-parameter family of self-adjoints
 extentions of $A$ in $\Sigma_{1}$.

For the inner portion of the bubble, the procedure is similar, but now with 
$r$ taking values between $-\infty$ and $r_{0}$. Again, there is a one-parameter
 family of self-adjoint extensions of $A$ in $L^2(\Sigma_{2})$.

Hence, the spacetime (\ref{esfera}) remains quantum mechanically singular. 
The evolution of a wave packet is not uniquely determined by the initial
 state of the particle and a boundary condition need to be imposed
 (independently) on both sides of the spherical wall.

\subsection{Cylindrical Shells}

For spacetime described by (\ref{cilindro}), the appropriate Hilbert space is
\begin{equation}
H=L^2(\Sigma_{1},|r-b|dr d\varphi dz)\otimes L^2(\Sigma_{2},|r-a|dr d\varphi dz).
\end{equation}

As in the previous case, after separating variables 
$\psi=R(r)e^{im\varphi}e^{ikz}$, we have (for the radial portion outside the shell)
\begin{equation}
R_{m,k}''(r)+\frac{1}{(r-b)}R_{m,k}'(r)+\bigg[(\pm i - 
(k^2+M^2))-\frac{m^2}{(r-b)^2}\bigg]R_{m,k}(r)=0.
\label{cilindricas}
\end{equation}

As $r$ goes to infinity, the asymptotic behavior of $R(r)$ is
\begin{equation}
R(r)\sim \frac{1}{\sqrt{(r-b)}}(C_{1}e^{\alpha (r-b) }+C_{2}e^{-\alpha (r-b)})
\end{equation}
where
\begin{equation}
\alpha=
\frac{1}{\sqrt{2}}\bigg[(\sqrt{1+(M^2+k^2)^2}+M^2+k^2)^{1/2} 
\mp 
i (\sqrt{1+(M^2+k^2)^2}-M^2-k^2)^{1/2}\bigg].
\end{equation}

Again, $R(r)$ is square-integrable near $+\infty$ only if $C_{1}=0$, hence
\begin{equation}
R(r)\sim \frac{1}{\sqrt{(r-b)}}e^{-\alpha (r-b)}.
\end{equation}

For $r$ near $r_{0}$, both solutions are continuous (as in the bubble
 case), hence square-integrable. Then we find a solution in $L^2$ for
 each equation in (\ref{teste}). The deficiency indices
 are $n_{\pm}=1$, so 
there are infinitely many self-adjoint extensions of $A$ in this case too. 
This space remains singular when tested by quantum particles.


\section{Boundary Conditions in $r=r_{0}$}
\label{section 5}

On both spacetimes studied in the last section  the Klein-Gordon equation   
presents only one singular point, $r=+\infty$ in the outer portion 
and $r=-\infty$ in the inner one (we remind that $r=a,b$ are polar metric 
singularities). The boundary conditions on $r=r_{0}$, necessary to turn 
self-adjoint the spatial portion of the wave operator are simple as we shall see. 

 The radial portions of the Klein-Gordon operator are given by 
\begin{equation}
\begin{aligned}
&A_{r}=\frac{1}{(r-b)^2}\bigg\{-\frac{d}{dr}\bigg[(r-b)^2\frac{d}{dr}\bigg]
+l(l+1)+M^2(r-b)^2\bigg\} \quad \textrm{(bubble)} \\&
A_{r}=\frac{1}{(r-b)}\bigg\{-\frac{d}{dr}\bigg[(r-b)\frac{d}{dr}\bigg]+
\frac{m^2}{(r-b)}+\\&\hspace{34mm}+(k^2+M^2)(r-b)\bigg\}
 \quad \textrm{(cylindrical shell)}
\end{aligned}
\end{equation}   
in the Hilbert spaces
\begin{equation}
\begin{array}{ll}
L^2((r_{0},\infty),(r-b)^2dr) \quad \textrm{(bubble)} \\
L^2((r_{0},\infty),|r-b|dr) \quad \textrm{(cylindrical shell)}.
\end{array}
\end{equation}

On both cases, they have the general form
\begin{equation}
A_{r}=-\frac{1}{w(r)}\bigg\{\frac{d}{dr}\bigg[p(r)\frac{d}{dr}\bigg]+q(r)\bigg\}
\label{forma geral}
\end{equation}
in the Hilbert space $L^2(\Sigma,w(r)dr)$.

Because $r=r_{0}$ is not a singular point of (\ref{forma geral}), 
the self-adjoint extensions are found extending the domain of $A$ to
 functions in $L^2$ satisfying \cite{richtmyer},
\begin{equation}
f(r_{0})\cos\alpha+p(r_{0})f'(r_{0})\sin\alpha=0
\end{equation}
with $\alpha\in \mathbb{R}$.

So, the boundary conditions in our case are:\\

{\it{\bf External Side}}\\

$\bullet$ {\bf Bubble}
\begin{equation}
R(r_{0})\cos \alpha+(r_{0}-b)^2R'(r_{0})\sin \alpha=0,
\label{cond}
\end{equation}

$\bullet$ {\bf Cylindrical Shell}
\begin{equation}
R(r_{0})\cos \alpha+(r_{0}-b)R'(r_{0})\sin \alpha=0,
\end{equation}
$\alpha \in \mathbb{R}$.\\

{\it{\bf Internal Side}}

$\bullet$ {\bf Bubble}
\begin{equation}
R(r_{0})\cos \beta+(r_{0}-a)^2R'(r_{0})\sin \beta=0,
\end{equation}

$\bullet$ {\bf Cylindrical Shell}
\begin{equation}
R(r_{0})\cos \beta+(r_{0}-a)R'(r_{0})\sin \beta=0,
\end{equation}
$\beta \in \mathbb{R}$.


\section{Massless Particle Outside the Spherical Wall}
\label{section 6}

To illustrate results of sections \ref{section 4} and \ref{section 5}, we will
 discuss a simple example,   a massless particle in the outer side of the
 bubble. The Klein-Gordon equation can be solved exactly by separation of 
variables $\Psi=R(r)Y_{l}^{m}(\theta,\varphi)e^{-i\omega t}$ 
(positive-frequency solution). The general solution of the radial part is
\begin{equation}
R_{l,m}(r)=A_{l,m}j_{l}(\omega (r-b))+B_{l,m} \eta_{l}(\omega (r-b)),
\label{solucao esferica}
\end{equation}
where $j_{l}$ and $\eta_{l}$ are the spherical Bessel and Neumann functions, 
respectively. Because neither of them are square-integrable near infinity, there
 is only one linear combination of these functions which are square-integrable 
\cite{richtmyer}, apart from a multiplicative constant. One of the constants 
in (\ref{solucao esferica}) was eliminated and we have a set
 of square-integrable functions at  $+\infty$, $\{F_{l,m}\}$, so that
\begin{equation}
R_{l,m}=C_{l,m}F_{l,m}(\omega (r-b)).
\end{equation}

We pick the parameter $\alpha=0$, which corresponds to the boundary condition 
$R(r_{0})=0$ (Dirichlet boundary condition). This condition imposes a 
quantization of the allowed energies $\{\omega_{l,n}\}_{n \in \mathbb{N}}$, 
where $\omega_{l,n}(r_0-b)$ are the zeros of $F_{l,m}$. We then have a 
complete set of eigenfunction $\{F_{l,m,n}\}$ so that a general solution 
of Klein-Gordon equation can be written by a superposition of the allowed 
modes as
\begin{equation}
\Psi(\vec{r},t)=\sum_{n=1}^{\infty}{\sum_{l=0}^{\infty}{\sum_{m=
-l}^{l}C_{l,m}F_{l,m,n}(r)Y_{l}^{m}(\theta,\varphi)e^{-i\omega_{n}t}}},
\end{equation}
resting only one constant to be determined. In this way, the knowledge of the 
 initial packet  is sufficient to determine the quantum dynamic of the particle.
 Note that the choice of one boundary conditions given in (\ref{cond}) turned
 our operator into self-adjoint, so that the evolution of the particle
 becomes uniquely determined by the initial wave packet.


\section{Conclusions}
\label{section 7}

Two singular spacetimes that represent  defects on a hypersurface were tested
 using  quantum mechanics. We  find that  they remain singular when probed by 
waves. They had to be partitioned in two independent parts, so the splitting 
of the Hilbert space in a tensorial product was necessary. Due to the fact
 that the classical singular points ($r-r_{0}=0$) are ordinary points of
 Klein-Gordon equation  the boundary conditions  necessary to turn
 self-adjoint the spatial portion of the wave operator  can be 
 easily found. Two real parameters are necessary, because each 
side of the wall behaves independently. 

A wide class of spacetimes with singularities in a $2$-dimensional
 submanifold can be found in references \cite{letelier2} and \cite{letelier1}.
 They  represent open and closed shells of different shapes, like parabolic 
and toroidal shells. These spacetimes are not as simple as   the ones studied 
in the present work, but they are still static and,
  in principle, can be treated, in a similar way.

We do not foresee special difficulties to study  Maxwell and Dirac 
fields for bubbles and cylindrical shells.

\acknowledgements

We thank CNPq for financial  support and  P.S.L. also thanks  FAPESP.


\end{document}